\documentclass[lettersize,journal,onecolumn]{IEEEtran}
\usepackage{setspace}
\usepackage{amsmath,amsfonts,amsthm,amssymb}
\usepackage{algorithmic}
\usepackage{algorithm}
\usepackage{array}
\usepackage[caption=false,font=normalsize,labelfont=sf,textfont=sf]{subfig}
\usepackage{textcomp}
\usepackage{stfloats}
\usepackage{url}
\usepackage{verbatim}
\usepackage{graphicx}
\usepackage{cite}
\usepackage{color}
\usepackage{booktabs}
\usepackage{enumitem}

\begin{document}

\title{Fundamental Limits of Joint Target Detection and Parameter Estimation — Characterizing Mixed-State Sensing Limits via Posterior Entropy Volume}

\author{Dazhuan Xu, Nan Wang and Han Zhang

\thanks{This work was supported by the National Natural Science Foundation of China under Grants 62271254, Grants 62601551, and Grants 662401286. This work was supported by the Fundamental Research Funds for the Universities of Liaoning Province (Grant No. LJ212610143055). This work was supported by the Natural Science Research Start-up Foundation of Recruiting Talents of Nanjing University of Posts and Telecommunications (Grant No. NY224007) (Corresponding author: Dazhuan Xu, e-mail: xudazhuan@nuaa.edu.cn)}
\thanks{Dazhuan Xu is with College of Electronic and Information Engineering, Nanjing University of Aeronautics and Astronautics, Nanjing, China and Purple Mountain Laboratories, Nanjing, China; Nan Wang is with College of Electronic and Information Engineering, Shenyang Aerospace University, Shenyang, China; Han Zhang is with School of Communications and Information Engineering, Nanjing University of Posts and Telecommunications, Nanjing, China. (e-mail: n\_wang@nuaa.edu.cn)}}

\markboth{Journal of \LaTeX\ Class Files,~Vol.~14, No.~8, August~2021}%
{Shell \MakeLowercase{\textit{et al.}}: A Sample Article Using IEEEtran.cls for IEEE Journals}


\maketitle
\begin{abstract}
The development of integrated sensing and communication calls for a unified theoretical foundation for sensing. This paper models target-presence patterns and continuous physical parameters as a mixed discrete–continuous state $\Xi$ on a branched reference measure, and treats posterior entropy volume and joint mutual information as two complementary representations of the same limit. Entropy volume carries physical units, can be compared with engineering scales such as resolution cells, and remains meaningful when the number of active targets varies; mutual information is dimensionless, invariant to coordinates and units, and additive through the chain rule. We define entropy number, entropy volume, and mixed entropy volume for discrete, continuous, and mixed states, respectively. The maximum-entropy principle for the uniform distribution on a support of fixed measure explains the measure-theoretic meaning of the exponential entropy scale, while the main limit follows from a mixed asymptotic equipartition property and a posterior probability–volume inequality through conditional typical sets. For asymptotically reliable high-probability sensing regions, the minimum achievable first-order posterior mixed entropy volume equals the prior mixed entropy volume multiplied by $2^{-I(\Xi;Y)}$, where $I(\Xi;Y)=I(V;Y)+I(X_V;Y\mid V)$. Thus, detection and estimation contributions multiply in the volume domain and add in the bit domain, and one sensing bit halves the posterior effective measure. We further prove that posterior-preserving cascades attain the direct-inference limit, while arbitrary intermediate compression incurs the exact information loss $I(\Xi;Y\mid Z)$. Numerical results for a single-target presence–range model illustrate the information composition, posterior entropy-volume contraction, and cascade-interface loss.
\end{abstract}

\section{Introduction}

Shannon information theory established for communication a complete framework built on probabilistic models, entropy and mutual information, capacity, and coding theorems \cite{ref1, ref2}. Against this background, integrated sensing and communication (ISAC) is becoming an important direction for next-generation wireless systems, and recent surveys have systematically reviewed sensing and communication performance metrics and their tradeoffs \cite{ref3, ref4}. By contrast, sensing has long been evaluated mainly through task-dependent quantities such as detection probability, false-alarm probability, mean-square error, Fisher information, the Cramér--Rao bound, and resolution. Information-theoretic studies of ISAC are by no means absent: prior work has established capacity--distortion tradeoffs for joint communication and state sensing \cite{ref5}, \cite{ref6} and has proposed frameworks that place sensing mutual information alongside communication mutual information \cite{ref7}. However, formulations constrained by average distortion do not naturally describe sensing criteria such as detection probability, a limitation already noted in the literature \cite{ref8}; more recent capacity--distortion analysis under logarithmic loss \cite{ref9} still considers states on a fixed alphabet. The problem addressed here is therefore to establish, for joint sensing with both discrete presence states and continuous physical parameters, a posterior-volume limit theorem for the complete mixed posterior, with both achievability and a converse.

Information theory has a long history in radar sensing. Woodward discussed radar delay measurement and detection from the viewpoint of posterior information gain \cite{ref10}, and Bell used mutual information for radar waveform design \cite{ref11}. Building on this line of work, the authors have developed frameworks of spatial information theory and sensing information theory \cite{ref12}, \cite{ref13}, and have studied target-detection information \cite{ref14}, range information \cite{ref15}, array spatial and directional information \cite{ref16}, joint range--Doppler limits \cite{ref17}, joint range--direction performance bounds \cite{ref18}, joint entropy error \cite{ref19}, and information bounds for compressive delay and compressive array sensing \cite{ref20}, \cite{ref21}. Preprints \cite{ref22, ref23, ref24} further considered fundamental limits for time-delay estimation, target detection, and joint detection--estimation in single-target settings. These works constitute the direct technical antecedents of the present paper, but they primarily concern purely discrete states, fixed-dimensional continuous states, or specific single-target radar models. The contribution here is not to restate a mutual-information chain rule, but to elevate entropy number, entropy volume, and mixed entropy volume into a unified uncertainty measure; to establish a mixed AEP and a high-probability posterior-volume converse on a general branched reference measure; and to reduce losslessness of a cascade rigorously to sufficiency of its intermediate representation for the complete mixed state.

Classical statistical inference commonly separates inference about an unknown state into parameter estimation and hypothesis testing \cite{ref25}. Target sensing, however, naturally contains both types of uncertainty: a system must determine whether targets exist and which targets are present, while also inferring continuous parameters such as range, direction, and Doppler for the targets that do exist. Traditional joint detection--estimation can jointly design detection and estimation rules within a Bayesian decision framework and optimize one-step or two-step decision procedures \cite{ref26}. Such methods answer how to make an optimal decision once losses and constraints have been specified. This paper addresses a more fundamental question: without prescribing a particular decision loss, how far can the observation itself compress the posterior uncertainty of the complete mixed state? The connection between mutual information and estimation error is also well known; for Gaussian channels, the I--MMSE relation links the derivative of mutual information with respect to signal-to-noise ratio to the optimal mean-square error \cite{ref27}. That relation, however, connects information to point-estimation risk under a particular observation model. Here the primary object is the complete posterior and its high-probability sensing region.

Our basic viewpoint is that the fundamental physical object for sensing performance is not the distance between a point estimate and the truth, but the reference measure of the high-probability posterior region in which the physical state may still lie after the observation is obtained. Continuous states correspond to Lebesgue measure, discrete states to counting measure, and mixed states to a branched reference measure composed of the two; entropy, conditional entropy, and information density on general alphabets can all be defined relative to a prescribed reference measure \cite{ref28}. The reference measure is part of the physical model, on the same footing as the prior and the sensing channel, rather than an arbitrarily chosen coordinate convention: for example, radar range is measured in meters and Doppler in hertz, as specified by the measurement system itself. More importantly, entropy volume is not merely a post hoc interpretation supplied by typical sets. A uniform distribution has maximum entropy on a support of fixed finite measure, so any distribution with a given entropy must have support measure no smaller than the corresponding exponential entropy scale. The mixed AEP then lifts this static maximum-entropy relation to a first-order achievable volume limit for high-probability posterior regions. The volume viewpoint is particularly natural for joint estimation of multiple physical quantities: for a range--Doppler state, the two coordinates have units of meters and hertz, and the joint posterior region has the natural reference scale m$\cdot$Hz, without artificially adding quantities in m$^2$ and Hz$^2$ through ad hoc weights.

Entropy volume and mutual information are in one-to-one correspondence through M\_H=2\^{}\ensuremath{\mathcal{H}} and are therefore two coordinates of the same mathematical limit. We do not claim that either is intrinsically more fundamental; instead, we make their roles explicit. Entropy volume carries physical units and depends on the reference measure, so it can be compared directly with engineering scales such as resolution cells and unambiguous windows, and it remains well defined when the number of active targets and hence the continuous dimension vary across branches. Mutual information is dimensionless, invariant to the choice of coordinates and units, directly comparable with information on the communication side, and additive through the chain rule. The fact that detection and estimation contributions multiply in the volume domain but add in the bit domain is precisely a manifestation of this division of roles. Accordingly, the paper uses volume to state physical conclusions and information to state additive structure; neither representation subsumes the other.

The main contributions are as follows. 1) We establish a mixed discrete--continuous sensing model on a finite-order branched reference measure and introduce a unified system of entropy number, entropy volume, and mixed entropy volume. 2) We connect the uniform maximum-entropy principle on a support of fixed measure with the mixed AEP and prove that the exponential of conditional entropy is the minimum achievable first-order measure scale of a high-probability posterior region. This yields a posterior entropy-volume characterization of the fundamental limit of joint detection--estimation and interprets I(\ensuremath{\Xi};Y) as the dimensionless logarithmic contraction rate of that volume. 3) We establish a losslessness theorem for posterior-preserving cascades and an exact information-loss identity for arbitrary intermediate compression, and show that pure detection and pure estimation arise as two boundary degenerations of the same mixed-state theorem.

\section{General Sensing-System Model on a Mixed State Space}

\subsection{General Sensing Channel}

Let the unknown physical state be \(\Xi \in S\) and the observation be \(Y\in\mathcal{Y}\). The sensing problem is jointly specified by the state prior \(P_{\Xi}\) and the conditional probability kernel \(P_{Y|\Xi}\). Unlike Shannon capacity, which maximizes over the input distribution \cite{ref1}, \cite{ref2}, we do not optimize \(P_{\Xi}\): the prior is part of the physical uncertainty of the environment rather than a freely designable channel-input distribution. We therefore use the term sensing information limit rather than sensing capacity. Capacity--distortion formulations that jointly optimize communication and sensing are studied in \cite{ref5}, \cite{ref6}; here we study the compression of state uncertainty produced by the sensing channel under a fixed physical prior. The sensing process is written as
\begin{equation}
	\label{1}
    \Xi\  \rightarrow \ Y\  \rightarrow \ p(\Xi \mid Y).
\end{equation}

The posterior \(p(\Xi \mid Y)\) is the fundamental state of knowledge obtained by the sensing system after observing the data. MAP estimation, MMSE estimation, hard decisions, confidence regions, and other point outputs are task-dependent decisions or compressions applied after the posterior is formed \cite{ref25}. We therefore separate posterior-level limits from the performance of specific estimators and detectors.

\subsection{Finite-Order Multi-Target Branched State}

Consider a scene with \(K\) potential target branches. Let \(V_{k} \in \{ 0,1\}\) indicate whether branch \(k\) is active, and write \(V = \left( V_{1},\ldots,V_{K} \right)\). If \(V_{k} = 1\), the target has a continuous physical parameter \(X_k\in\mathcal{X}\subset\mathbb{R}^{d}\); if \(V_{k} = 0\), then \(X_{k}\) does not correspond to an actual physical degree of freedom. Define the active set \(\mathcal{I}(V) = \{ k:V_{k} = 1\}\) and the active continuous state \(X_V=(X_k:k\in\mathcal{I}(V))\). The scene state is
\begin{equation}
	\label{2}
    \Xi = \left( V,\ X_{V} \right).
\end{equation}

The corresponding state space has a continuous dimension determined by the discrete presence pattern:

\begin{equation}
	\label{3}
	S = \bigsqcup_{v \in \{ 0,1\}^{K}}\{ v\} \times \mathcal{X}^{|v|}.
\end{equation}

The reference measure is counting measure over the discrete branches and Lebesgue measure over the active continuous coordinates. This construction expresses, at the measure-theoretic level, the physical fact that a parameter has no meaning when its target is absent. We assume finite \(K\) and do not address random finite sets, unlabeled permutations, or data-association problems.

\subsection{General Matrix Sensing-System Model}

To connect the abstract sensing channel with practical radar and array systems, consider \(L\) snapshots, pulses, or equivalent observation dimensions. For the active targets, construct a physical-mode matrix \(U\left( X_{V} \right)\), and let \(S_{V}\) denote the target scattering or response matrix. A general model is

\begin{equation}
	\label{4}
	Y = U\left( X_{V} \right)\, S_{V} + W,
\end{equation}

where \(Y \in \mathbb{C}^{N \times L}\) is the observation matrix, the columns of \(U\left( X_{V} \right)\) are the physical-mode vectors \(u\left( X_{k} \right)\) of the active targets, \(S_{V} \in \mathbb{C}^{|V| \times L}\), and \(W\) represents noise, interference, or unmodeled disturbances. This form makes explicit the chain from physical state to physical mode, target response, and observation. The theorems below require only the conditional probability kernel \(P_{Y|\Xi}\) and do not depend on linearity or Gaussian noise.

\subsection{Two Special Cases: Ranging and Array Direction Finding}

Range sensing, array direction finding, and joint multiparameter estimation have led in the authors' previous work to range information, spatial and directional information, and joint-parameter performance limits \cite{ref15}--\cite{ref19}. Here they serve only as physical instances of the general sensing mode \(u(x)\); the theorems are not tied to a particular waveform, array manifold, or scattering model. For ranging, let \(X_{k} = \tau_{k}\), with \(u\left( \tau_{k} \right)\) formed from delayed replicas of the transmitted waveform. For array direction finding, let \(X_{k} = \theta_{k}\), with \(u\left( X_{k} \right) = a\left( X_{k} \right)\) equal to the array manifold. Range--direction, range--Doppler, and range--direction--Doppler sensing require only an enlarged continuous state coordinate and physical-mode map; the information-theoretic results remain unchanged.

\section{Entropy Number, Entropy Volume, and Joint Sensing Information}

\subsection{Basic Principle of the Unified Measure}

We take the exponential form of entropy as the basic geometric object. For a discrete state, exponential entropy gives an effective number of candidates; for a continuous state, exponential differential entropy gives an exponential volume scale relative to the reference measure; and for a mixed state, it gives an effective measure scale on the branched reference measure \cite{ref28}. This is a measure-theoretic interpretation, not a claim that \(2^{h}\) is an absolute geometric volume invariant under arbitrary coordinate transformations. The entropy volume of a continuous state is a physical quantity that depends on the reference measure, whereas mutual information is the difference between prior and posterior log-volumes, so the reference scale cancels. The term physical bit refers to the fact that the log-volume representation of differential entropy inherits the physical reference scale; sensing information itself remains an ordinary number of bits measuring logarithmic compression of physical uncertainty volume. The reference measure is specified by the physical measurement system and is part of the model rather than a freely chosen coordinate convention. Once the model is fixed, entropy volume is well defined. The fact that entropy volume depends on the reference measure whereas mutual information does not is not a hierarchy of importance, but the source of their complementary roles.

\subsection{Entropy Number}

\noindent\textbf{Definition 1 (Entropy Number).} For a discrete state \(V\), define the posterior entropy number by

\begin{equation}
	\label{5}
	N_{H}(V \mid Y) \triangleq 2^{H(V \mid Y)}.
\end{equation}

\(N_{H}\) is not the literal cardinality of the posterior support, but the effective number of candidates determined by Shannon entropy. For a binary target-presence state, \(1 \leq N_{H} \leq 2\). If the observation provides no information about target presence, \(N_{H}\) remains near the prior effective candidate count; if the posterior concentrates on a single hypothesis, \(N_{H} \rightarrow 1\). The quantity \(N_{H}\) is the same as \(N_{hyp}\) in the authors' earlier work; throughout this paper we use the notation \(N_{H}\).

\subsection{Entropy Volume}

\noindent\textbf{Definition 2 (Posterior Entropy Volume).} For a \(d\)-dimensional continuous physical state \(X\), under a specified reference coordinate system and Lebesgue measure, define the posterior entropy volume by

\begin{equation}
	\label{6}
	V_{H}(X \mid Y) \triangleq 2^{h(X \mid Y)}.
\end{equation}

\(V_{H}\) is the single-sample exponential volume scale of the conditional typical set and is the basic physical measure of continuous sensing uncertainty. If \(X\) is range, delay, or Doppler, \(V_{H}\) has the corresponding reference scale m, s, or Hz. If \(X = \left( R,f_{D} \right)\), the reference scale of \(V_{H}\) is m$\cdot$Hz.

\noindent\textbf{Definition 3 (Gaussian-Calibrated Entropy Volume).} Within a continuous branch of fixed dimension \(d\), define

\begin{equation}
	\label{7}
	V_{E}(X \mid Y) \triangleq (2\pi e)^{- d/2}\, 2^{h(X \mid Y)}.
\end{equation}

If \(X \mid Y\) is a \(d\)-dimensional Gaussian distribution \(\mathcal{N}(\mu,C)\), then \(V_{E} = \sqrt{\det C}\), and the generalized entropy error satisfies \(GEE = V_{E}^{2} = detC\). This calibration is consistent with the classical connection between entropy power and estimation \cite{ref2}, \cite{ref27} and links to the authors' previous joint entropy-error bounds \cite{ref18}, \cite{ref19}. Gaussian normalization is used only within a fixed continuous dimension. In a mixed state with a varying number of active targets, different branches have different dimensions and different physical volume elements; the unified theory must therefore be based on \(2^{\mathcal{H}}\) with respect to the branched reference measure rather than applying the same \((2\pi e)^{- d/2}\) factor to every branch.

\subsection{Mixed Entropy Volume}

For the mixed state \(\Xi = \left( V,X_{V} \right)\), define the mixed conditional entropy relative to the branched reference measure \(\mu\) as

\begin{equation}
	\label{8}
	\mathcal{H}(\Xi \mid Y) \triangleq H(V \mid Y) + h\left( X_{V} \mid V,Y \right).
\end{equation}

\textbf{Definition 4 (Mixed Entropy Volume).}

\begin{equation}
	\label{9}
	M_{H}(\Xi \mid Y) \triangleq 2^{\mathcal{H}(\Xi \mid Y)}.
\end{equation}

The chain decomposition of mixed entropy gives the multiplicative structure

\begin{equation}
	\label{10}
	M_{H}(\Xi \mid Y) = N_{H}(V \mid Y) \cdot V_{H}\left( X_{V} \mid V,Y \right),
\end{equation}

where \(V_{H}\left( X_{V} \mid V,Y \right) \triangleq 2^{h\left( X_{V} \mid V,Y \right)}\). The discrete component is measured by an entropy number, the continuous component by a conditional entropy volume, and the joint state by their product: information adds in the logarithmic domain, whereas uncertainty multiplies in the physical-measure domain. Mixed entropy volume is always understood as an effective measure scale relative to the prescribed branched reference measure, not as an ordinary geometric volume in a fixed-dimensional Euclidean space.

\noindent\textit{Note 1.} Within a continuous branch of fixed dimension \(d\), \(M_{H} = (2\pi e)^{d/2}ED_{J}\), where \(ED_{J}\) is the joint entropy deviation used in \cite{ref19} and in the monograph \cite{ref13}; \(V_{E}^{2} = GEE\); and \(N_{H} = N_{hyp}\). We take \(2^{\mathcal{H}}\) as the fundamental quantity because the factor \((2\pi e)^{- d/2}\) cannot be applied uniformly across variable-dimensional branches. On their common domain of applicability, the two sets of measures differ only by explicit constants and the corresponding results can be translated between them.

\subsection{Maximum-Entropy Characterization of Entropy Volume}

The basic measure-theoretic foundation of entropy volume is that the uniform distribution maximizes entropy on a support of fixed measure. Let a probability distribution \(P\) be absolutely continuous with respect to a reference measure \(\mu\) and supported on a measurable set \(A\) with \(0<\mu(A)<\infty\). Let \(U_A\) denote the distribution that is uniform on \(A\) relative to \(\mu\). From \(D(P\|U_A)\geq 0\), we obtain
\[
\mathcal{H}_{\mu}(P)\leq \log_2\mu(A),
\]
and hence
\[
\mu(A)\geq 2^{\mathcal{H}_{\mu}(P)}.
\]
Equality holds if and only if P=U\_A \ensuremath{\mu}-a.e. Thus \(2^{\mathcal{H}_{\mu}(P)}\) is not an arbitrarily assigned geometric label for entropy; it is the minimum support-measure scale compatible with the given entropy. In a discrete space it reduces to an effective candidate count, in a continuous space to an effective volume, and in a branched mixed space to an effective mixed-measure scale relative to a counting$\times$Lebesgue-type reference measure. The statement depends only on the reference measure and not on whether the state space is purely discrete, purely continuous, or mixed.

A distinction must be made between full support and a high-probability posterior region. The maximum-entropy principle directly constrains a support carrying all probability mass. A practical sensing region need only contain the true state with probability tending to one, so the static inequality above alone cannot establish the main converse. Sections IV and V use the mixed AEP to show that the conditional posterior has exponential density scale \(2^{-m\mathcal{H}(\Xi\mid Y)}\) on a typical set, and then use a probability--volume inequality to lift the maximum-entropy idea of support volume into an asymptotically achievable limit for high-probability posterior entropy volume. In short, the maximum-entropy principle explains why exponential entropy is a volume, while the mixed AEP proves why that volume is the sensing limit.

Note (Roles of Three Cases). To avoid ambiguity in the word volume, we distinguish three cases. First, for a single observation (m=1) when the region is required to carry the entire posterior probability, the maximum-entropy inequality directly yields \(\mu(A)\geq 2^{\mathcal{H}(\Xi\mid Y=y)}\), a volume bound in the single-scene state space. Second, for an m-fold extension when the region need only contain the true state with probability tending to one, the mixed AEP and posterior probability--volume inequality in Sections IV and V yield achievability and a converse. Here \(M_m=\mu^m(R_m)^{1/m}\) is the per-state geometric mean of a region measure in S\^{}m, not a single region in the one-scene state space; it should not be visualized as an ellipse in a (range, Doppler) plane. Third, for a single observation with only a high-probability containment requirement, the corresponding volume bound does not hold. This is precisely why the main theorem requires the mixed AEP and cannot follow from the maximum-entropy principle alone.

\subsection{Joint Sensing Information and the Entropy-Volume Contraction Law}

Define the joint sensing information as
\begin{equation}
	\label{11}
I_{J} \triangleq I(\Xi;Y)\mathcal{= H}(\Xi)\mathcal{- H}(\Xi \mid Y).
\end{equation}

By the chain rule for mutual information,

\begin{equation}
	\label{12}
	I_{J} = I(V;Y) + I\left( X_{V};Y \mid V \right) \triangleq I_{D} + I_{E}.
\end{equation}

Equation (12) is operationally significant for joint sensing: detection information and conditional estimation information add directly on the same information scale, without artificial normalization or weighting. For a single-target branched state, the continuous-information term contributes only on the target-present branch and therefore naturally carries the branch probability weight. Exponentiating the entropy difference gives the posterior entropy-volume contraction law

\begin{equation}
	\label{13}
	M_{H}(\Xi \mid Y) = M_{H}(\Xi) \cdot 2^{- I(\Xi;Y)}, 2^{\, I(\Xi;Y)} = \frac{M_{H}(\Xi)}{M_{H}(\Xi \mid Y)}.
\end{equation}

One sensing bit therefore has the following operational physical meaning: under the same reference measure, the posterior effective uncertainty scale is reduced by one half relative to the prior. For a one-dimensional continuous state, the effective length is halved; in two dimensions the effective area is halved; in d dimensions the effective volume is halved; and for a discrete state the effective candidate count is halved. Differential entropy may retain reference labels such as bit\_m, bit\_s, bit\_Hz, or bit\_\{m$\cdot$Hz\} to indicate the underlying physical log-volume scale. Mutual information is the difference between prior and posterior log-volumes in the same state space, so the reference scale cancels and the result remains an ordinary number of bits.

\section{From Maximum Entropy to the Mixed AEP: Geometry of Posterior Entropy Volume}

\subsection{Probability Space, Reference Measure, and Regularity Conditions}

Let \((\Xi,Y)\) have joint distribution \(P_{\Xi Y}\). The mixed physical state \(\Xi\) is defined on the measurable space \(\left( S\mathcal{,S} \right)\) and has density \(p_{\Xi}\) with respect to a \(\sigma\)-finite branched reference measure \(\mu\); the joint distribution has density \(p_{\Xi Y}\) with respect to \(\mu \times \nu\), and the conditional density \(p(\xi \mid y)\) is defined by the Radon--Nikodym derivative. Consider an \(m\)-fold memoryless extension in which \(\left( \Xi_{i},Y_{i} \right)_{i = 1}^{m}\) are i.i.d. according to \(P_{\Xi Y}\).

\noindent\textbf{Assumption A (Integrability).} The relevant log-densities have finite absolute first moments:

\begin{equation}
	\label{14}
	\begin{array}{c}
		\begin{aligned}
		&\mathbb{E}\!\left[\left|\log_2 p_{\Xi}(\Xi)\right|\right]<\infty,\\
		&\mathbb{E}\!\left[\left|\log_2 p_{\Xi Y}(\Xi,Y)\right|\right]<\infty,\\
		&\mathbb{E}\!\left[\left|\log_2 p_Y(Y)\right|\right]<\infty .
		\end{aligned}
	\end{array}
\end{equation}

Under Assumption A, quantities such as \(\mathcal{H}(\Xi)\), \(\mathcal{H}(\Xi \mid Y)\), and \(h(Y)\) are finite, and the weak law of large numbers can be used to establish mixed asymptotic equipartition. For a finite-order branched state, the assumption can be checked branch by branch. The weak law, information density, and conditional typical sets used in this section are standard tools \cite{ref2}, \cite{ref28}. For general stationary ergodic processes with densities, Barron\textquotesingle s generalized Shannon--McMillan--Breiman theorem \cite{ref29} gives a stronger result; the i.i.d. finite-branch setting considered here does not require it.

\noindent\textit{Notation.} For a random sequence \(\{ A_{m}\}\) and a constant \(a\), write \({p - liminf}_{m \rightarrow \infty}A_{m} \geq a\) if \(Pr\{ A_{m} < a - \zeta\} \rightarrow 0\) for every \(\zeta > 0\). The quantity \(p - limsup\) is defined analogously.

\subsection{Mixed Asymptotic Equipartition}

Define the mixed weakly typical set
\begin{equation}
	\label{15}
    A_{\varepsilon}^{(m)}(\Xi) \triangleq \{\xi^{m}:\ | - \frac{1}{m}\log_{2}p\left( \xi^{m} \right)\mathcal{- H}(\Xi)\left| < \varepsilon \right\}.
\end{equation}

\noindent\textbf{Lemma 1 (Mixed Asymptotic Equipartition).} Under Assumption A, for every \(\varepsilon > 0\):

\begin{enumerate}[
	label=(\roman*),
	align=left,
	labelwidth=1.6em,
	labelsep=0.4em,
	leftmargin=2.0em
	]
	\item
	$\Pr\!\left\{
	\Xi^m\in A_{\varepsilon}^{(m)}(\Xi)
	\right\}\to 1$;
	
	\item
	for every
	$\xi^m\in A_{\varepsilon}^{(m)}(\Xi)$,
	\begin{equation}
		2^{-m[\mathcal{H}(\Xi)+\varepsilon]}
		\leq
		p(\xi^m)
		\leq
		2^{-m[\mathcal{H}(\Xi)-\varepsilon]}.
		\label{eq:mixed_aep_density}
	\end{equation}
	
	\item
	for all sufficiently large $m$,
\end{enumerate}

\begin{equation}
	\label{17}
	(1 - \varepsilon)\, 2^{m\left\lbrack \mathcal{H}(\Xi) - \varepsilon \right\rbrack}\  \leq \ \mu^{m}\left( A_{\varepsilon}^{(m)}(\Xi) \right)\  \leq \ 2^{m\left\lbrack \mathcal{H}(\Xi) + \varepsilon \right\rbrack}.
\end{equation}

\noindent\textit{Proof.} Let \(Z_{i} = - \log_{2}p_{\Xi}\left( \Xi_{i} \right)\). Under Assumption A, \(\{ Z_{i}\}\) are i.i.d. with \(\mathbb{E}[Z_i]=\mathcal{H}(\Xi)\). The weak law of large numbers gives \(\frac{1}{m}\sum_i Z_i\rightarrow\mathcal{H}(\Xi)\) in probability, proving (i). Part (ii) is a rewriting of (15). For (iii), integrate the lower bound in (16) over the typical set with respect to \(\mu^{m}\) and use that total probability is at most one to obtain the upper measure bound; integrate the upper density bound and combine it with (i) to obtain the lower measure bound. \hfill$\square$

The proof of Lemma 1 uses only the existence of a density with respect to \(\mu\) and integrability of its logarithm. It does not require \(\mu\) to be Lebesgue measure or counting measure, so the branched structure needs no special treatment.

\subsection{Conditional Atypical Probability}

The joint typical set \(A_{\varepsilon}^{(m)}(\Xi,Y)\) is defined by requiring the state, observation, and joint deviations to satisfy their respective typicality conditions simultaneously. For a given \(y^{m}\), define the conditional typical set by

\begin{equation}
	\label{18}
A_{\varepsilon}^{(m)}\left( \Xi \mid y^{m} \right) \triangleq \left\{ \xi^{m}:\ \left( \xi^{m},y^{m} \right) \in A_{\varepsilon}^{(m)}(\Xi,Y) \right\},
\end{equation}

Define the conditional atypical probability and its average as

\begin{equation}
	\label{19}
		\begin{array}{c}
		\begin{aligned}
				&q_{m}\left( y^{m} \right) \triangleq Pr\{\Xi^{m} \notin A_{\varepsilon}^{(m)}\left( \Xi \mid y^{m} \right)\,\left| \, Y^{m} = y^{m} \right\},\\
				&\delta_{m}\mathbb{\triangleq E}\left\lbrack q_{m}\left( Y^{m} \right) \right\rbrack.
		\end{aligned}
	\end{array}
\end{equation}

Applying Lemma 1 to the joint sequence gives \(\delta_{m} = Pr\{\left( \Xi^{m},Y^{m} \right) \notin A_{\varepsilon}^{(m)}(\Xi,Y)\} \rightarrow 0\). For any fixed \(a \in (0,1)\), Markov\textquotesingle s inequality yields

\begin{equation}
	\label{20}
\Pr\left\{ q_{m}\left( Y^{m} \right) > \delta_{m}^{\, a} \right\}\  \leq \ \delta_{m}^{\, 1 - a}\  \rightarrow \ 0,
\end{equation}

Thus, although the conditional atypical probability need not converge pointwise, it is uniformly small on a set of observations whose probability tends to one. Set \(a = 1/2\) and define

\begin{equation}
	\label{21}
	\begin{array}{c}
		\begin{aligned}
			&\mathcal{B}_{m} \triangleq \left\{ y^{m}:\ q_{m}\left( y^{m} \right) \leq \sqrt{\delta_{m}} \right\},\\
			&Pr\{ Y^{m} \in \mathcal{B}_{m}\}\  \geq \ 1 - \sqrt{\delta_{m}}.
		\end{aligned}
	\end{array}
\end{equation}

\subsection{Conditional Mixed Asymptotic Equipartition}

\noindent\textbf{Lemma 2 (Conditional Mixed Asymptotic Equipartition).}
Under Assumption A, there exists $c_{\varepsilon}$ such that
$c_{\varepsilon}\to 0$ as $\varepsilon\to 0$, and:

\begin{enumerate}[
	label=(\roman*),
	align=left,
	labelwidth=1.6em,
	labelsep=0.35em,
	leftmargin=1.95em
	]
	
	\item
	for every jointly typical pair
	$(\xi^m,y^m)\in A_{\varepsilon}^{(m)}(\Xi,Y)$,
	\begin{equation}
		2^{-m[\mathcal{H}(\Xi\mid Y)+c_{\varepsilon}]}
		\leq
		p(\xi^m\mid y^m)
		\leq
		2^{-m[\mathcal{H}(\Xi\mid Y)-c_{\varepsilon}]}.
		\label{eq:conditional_density_bound}
	\end{equation}
	
	\item
	for every $y^m$,
	\begin{equation}
		\mu^m\!\left(
		A_{\varepsilon}^{(m)}(\Xi\mid y^m)
		\right)
		\leq
		2^{m[\mathcal{H}(\Xi\mid Y)+c_{\varepsilon}]}.
		\label{eq:conditional_measure_upper}
	\end{equation}
	
	\item
	for $y^m\in\mathcal{B}_m$,
	\begin{equation}
		\mu^m\!\left(
		A_{\varepsilon}^{(m)}(\Xi\mid y^m)
		\right)
		\geq
		\left(1-\sqrt{\delta_m}\right)
		2^{m[\mathcal{H}(\Xi\mid Y)-c_{\varepsilon}]}.
		\label{eq:conditional_measure_lower}
	\end{equation}
	
\end{enumerate}

\noindent\textit{Proof.} (i) For a jointly typical pair, use the identity \(- \log_{2}p\left( \xi^{m} \mid y^{m} \right) = - \log_{2}p\left( \xi^{m},y^{m} \right) + \log_{2}p\left( y^{m} \right)\) and apply bounds of the form (16) separately to the joint density and the observation marginal. The total deviation is no larger than the sum of the two tolerances, denoted \(c_{\varepsilon}\). (ii) Integrate the lower density bound in (22) over \(A_{\varepsilon}^{(m)}\left( \Xi \mid y^{m} \right)\) with respect to \(\mu^{m}\) and use that the conditional probability is at most one. (iii) By (21), when \(y^{m} \in \mathcal{B}_{m}\) the conditional probability of this set is at least \(1 - \sqrt{\delta_{m}}\); integrating the upper density bound in (22) then gives the result. \hfill$\square$

The measure lower bound (24) holds only on the high-probability observation set \(\mathcal{B}_{m}\), and this restriction is essential. The density bound (22) and measure upper bound (23), by contrast, hold for all corresponding objects. From (17), (23), and (24), the ratio between the prior typical-set measure and the conditional typical-set measure on \(\mathcal{B}_{m}\) satisfies

\begin{equation}
	\label{25}
	\frac{\mu^{m}\left( A_{\varepsilon}^{(m)}(\Xi) \right)}{\mu^{m}\left( A_{\varepsilon}^{(m)}\left( \Xi \mid y^{m} \right) \right)}\  \doteq \ 2^{\, mI(\Xi;Y)}.
\end{equation}

The quantity constrained here is the effective mixed measure; no particular shape is imposed on the posterior region.

Entropy-volume interpretation. Equations (22) and (24) together show that, for high-probability observation sequences, the conditional posterior is approximately spread uniformly over the conditional typical set to first exponential order. The exponential order of the typical density is determined by the conditional entropy, while the exponential order of the typical-set measure is exactly the corresponding conditional entropy-volume scale. This is the asymptotic posterior counterpart of the maximum-entropy principle in Section III: to first exponential order, the conditional typical set realizes the entropy volume prescribed by the conditional entropy.

\subsection{Sensing Operators and Empirical Entropy Volume}

\noindent\textbf{Definition 5 (Sensing Operator).} Given \(Y^{m}\) and an internal random variable \(U\) (\(U\) is independent of \(\left( \Xi^{m},Y^{m} \right)\)), a sensing operator outputs a measurable region \(R_{m} = R_{m}\left( Y^{m},U \right) \subset S^{m}\) and may also output a representative sequence \({\widehat{\Xi}}^{m} \in R_{m}\). Define the failure probability, empirical sensing entropy, empirical sensing information, and empirical entropy volume by

\begin{equation}
	\label{26}
	\begin{array}{c}
		\begin{aligned}
			&P_{f}^{(m)} \triangleq Pr\{\Xi^{m} \notin R_{m}\},\\
			&\mathcal{H}_{m} \triangleq \frac{1}{m}\log_{2}\mu^{m}\left( R_{m} \right),\\
			&J_{m}\mathcal{\triangleq H}(\Xi) - \mathcal{H}_{m},
		\end{aligned}
	\end{array}
\end{equation}

\begin{equation}
	\label{27}
	M_{m} \triangleq \mu^{m}\left( R_{m} \right)^{1/m} = 2^{\mathcal{H}_{m}}.
\end{equation}

Taking the \(m\)th root converts the total mixed measure of the extended system back to an equivalent posterior volume scale per physical state. \(\mathcal{H}_{m}\), \(J_{m}\), and \(M_{m}\) are random variables. A sequence of sensing operators is called asymptotically reliable if \(P_{f}^{(m)} \rightarrow 0\).

\textbf{Definition 6 (Typical-Set Sensor). A sensing operator with}

\begin{equation}
	\label{28}
	R_{m} = A_{\varepsilon}^{(m)}\left( \Xi \mid Y^{m} \right),\quad\quad{\widehat{\Xi}}^{m} \sim p\left( \xi^{m} \mid Y^{m} \right)
\end{equation}

is called a typical-set sensor; its representative sequence is generated by posterior sampling.

\section{Posterior Entropy-Volume Limit of Joint Target Detection and Parameter Estimation}

The maximum-entropy principle in Section III gives the static relation ``entropy $\rightarrow$ minimum full-support measure,'' while the mixed AEP in Section IV gives exponential equipartition of the conditional posterior. This section further removes the requirement of carrying all probability mass by using a high-probability posterior probability--volume inequality. We first establish an achievability--converse theorem for posterior entropy volume and then express joint sensing information as the logarithm of the corresponding volume contraction. The static relation in Section III provides the measure-theoretic basis for the concept of entropy volume; the main proof chain consists of Lemmas 1--5 and does not rely on that static inequality as a premise.

\subsection{Achievability}

\textbf{Lemma 3 (Properties of the Typical-Set Sensor). Under Assumption A, the typical-set sensor satisfies:}

\begin{enumerate}[
	label=(\roman*),
	align=left,
	leftmargin=*,
	widest=iii
	]
	\item $P_f^{(m)}=\delta_m\to 0$;
	
	\item $\mathcal{H}_m\to\mathcal{H}(\Xi\mid Y)$ in probability,
	and hence
	$J_m\to I(\Xi;Y)$ and
	$M_m\to 2^{\mathcal{H}(\Xi\mid Y)}$
	in probability;
	
	\item the representative sequence has the same distribution as
	the true state sequence,
\end{enumerate}

\begin{equation}
	\label{29}
	\left( {\widehat{\Xi}}^{m},\ Y^{m} \right)\ \overset{d}{=}\ \left( \Xi^{m},\ Y^{m} \right).
\end{equation}

\noindent\textit{Proof.} (i) By (18), \(\Xi^{m} \notin A_{\varepsilon}^{(m)}\left( \Xi \mid Y^{m} \right)\) if and only if \(\left( \Xi^{m},Y^{m} \right) \notin A_{\varepsilon}^{(m)}(\Xi,Y)\). Hence \(P_{f}^{(m)} = \delta_{m}\), and Lemma 1 applied to the joint sequence gives \(\delta_{m} \rightarrow 0\).

(ii) For \(Y^{m} \in \mathcal{B}_{m}\), (23) and (24) give

\begin{equation}
	\label{30}
	|\mathcal{H}_{m}\mathcal{- H}(\Xi \mid Y)|\  \leq \ c_{\varepsilon} + \frac{1}{m}\log_{2}\frac{1}{1 - \sqrt{\delta_{m}}},
\end{equation}

Moreover, \(Pr\{ Y^{m} \in \mathcal{B}_{m}\} \geq 1 - \sqrt{\delta_{m}} \rightarrow 1\). First let \(m \rightarrow \infty\) and then \(\varepsilon \rightarrow 0\); this yields \(\mathcal{H}_m\rightarrow\mathcal{H}(\Xi\mid Y)\) in probability. The corresponding conclusions for \(J_{m}\) and \(M_{m}\) follow from (26), (27), and the continuous mapping theorem.

(iii) By (28), the conditional distribution of \({\widehat{\Xi}}^{m}\) given \(Y^{m}\) is the same as the true posterior distribution of \(\Xi^{m}\). Multiplying by the marginal distribution of \(Y^{m}\) yields (29). \hfill$\square$

Lemma 3(iii) shows that the output of the typical-set sensor and the state sequence generated by the physical world are indistinguishable in joint distribution. A method that selects a single point according to a fixed criterion generally does not have this property. Equation (29), however, does not imply that a randomized point output is superior to MAP or MMSE estimation under a specified finite-dimensional Bayesian risk \cite{ref25}. The achievable object in this paper is a high-probability sensing region together with a distribution-preserving representative sequence.

\subsection{Posterior Probability--Entropy-Volume Bound}

\noindent\textbf{Lemma 4 (Posterior Probability--Volume Bound).} For any measurable sensing region \(R_{m}\left( y^{m} \right)\), let \(s_{m}\left( y^{m} \right) \triangleq Pr\{\Xi^{m} \in R_{m}\left( y^{m} \right) \mid Y^{m} = y^{m}\}\). Then

\begin{equation}
	\label{31}
	s_{m}\left( y^{m} \right)\  \leq \ \mu^{m}\left( R_{m}\left( y^{m} \right) \right) \cdot 2^{- m\left\lbrack \mathcal{H}(\Xi \mid Y) - c_{\varepsilon} \right\rbrack}\  + \ q_{m}\left( y^{m} \right).
\end{equation}

\noindent\textit{Proof.} Fix \(y^{m}\) and write \(A_{m} = A_{\varepsilon}^{(m)}\left( \Xi \mid y^{m} \right)\). Decompose the success probability according to whether the state falls in the conditional typical set:

\begin{equation}
	\label{32}
		\begin{array}{c}
		\begin{aligned}
			&s_{m}\left( y^{m} \right) = \\
			&\int_{R_{m} \cap A_{m}}^{}p\left( \xi^{m} \mid y^{m} \right)\, d\mu^{m} + \int_{R_{m} \cap A_{m}^{c}}^{}p\left( \xi^{m} \mid y^{m} \right)\, d\mu^{m}.
		\end{aligned}
	\end{array}
\end{equation}

The second term is at most \(q_{m}\left( y^{m} \right)\); by the density upper bound in (22), the first term is at most \(\mu^{m}\left( R_{m}\left( y^{m} \right) \right)\, 2^{- m\left\lbrack \mathcal{H}(\Xi \mid Y) - c_{\varepsilon} \right\rbrack}\). \hfill$\square$

For a sensing operator with an internal random variable \(U\), apply Lemma 4 pointwise conditional on \(U\) and then average over \(U\); the conclusion is unchanged.

\subsection{Region Converse}

\textbf{Lemma 5 (Region Converse). For any asymptotically reliable sequence of sensing operators,}

\begin{equation}
	\label{32}
	\begin{array}{c}
		\begin{aligned}
			&\underset{m \rightarrow \infty}{p - liminf}\ \mathcal{H}_{m}\  \geq \ \mathcal{H}(\Xi \mid Y),\\
			&\underset{m \rightarrow \infty}{p - limsup}\ J_{m}\  \leq \ I(\Xi;Y).
		\end{aligned}
	\end{array}
\end{equation}

\noindent\textit{Proof.} Fix \(\varepsilon > 0\) and \(\eta > 0\), and define the event that the sensing region is too small by

\begin{equation}
	\label{34}
	B_{m}(\eta) \triangleq \left\{ \,\mathcal{H}_{m}\mathcal{< H}(\Xi \mid Y) - c_{\varepsilon} - \eta\, \right\}.
\end{equation}

On \(B_{m}(\eta)\), \(\mu^{m}\left( R_{m} \right) \leq 2^{m\left\lbrack \mathcal{H}(\Xi \mid Y) - c_{\varepsilon} - \eta \right\rbrack}\). Substituting this into (31) gives \(s_{m} \leq 2^{- m\eta} + q_{m}\). Taking expectation over \(Y^{m}\) and separating \(B_{m}(\eta)\) from its complement, on which the success probability is at most one, gives

\begin{equation}
	\label{35}
	1 - P_{f}^{(m)}\  \leq \ \Pr\left( B_{m}^{c}(\eta) \right) + 2^{- m\eta}\Pr\left( B_{m}(\eta) \right) + \delta_{m},
\end{equation}

Rearranging yields

\begin{equation}
	\label{36}
	\Pr\left( B_{m}(\eta) \right)\  \leq \ \frac{P_{f}^{(m)} + \delta_{m}}{1 - 2^{- m\eta}}.
\end{equation}

Because \(P_{f}^{(m)} \rightarrow 0\) and \(\delta_{m} \rightarrow 0\), for every fixed \(\varepsilon,\eta > 0\) we have \(\Pr\left( B_{m}(\eta) \right) \rightarrow 0\). Let \(m \rightarrow \infty\) first, then let \(\varepsilon \rightarrow 0\) and \(\eta \rightarrow 0\); this gives the first relation in (33). The second follows from \(J_m=\mathcal{H}(\Xi)-\mathcal{H}_m\). \hfill$\square$

Lemma 5 directly constrains the random sensing region itself. This is necessary: a lower bound on \(\log\mathbb{E}\left\lbrack \mu^{m}\left( R_{m} \right) \right\rbrack\) is insufficient to constrain \(J_{m}\), because Jensen\textquotesingle s inequality gives \(\log\mathbb{E}\left\lbrack \mu^{m}\left( R_{m} \right) \right\rbrack\mathbb{\geq E}\log\mu^{m}\left( R_{m} \right)\) in the opposite direction from what is needed. Events of very small probability but extremely large measure can inflate the mean measure without affecting the random empirical entropy.

\subsection{Posterior Entropy-Volume Limit: Primary Form}

\noindent\textbf{Theorem 1 (Posterior Entropy-Volume Limit).} Under Assumption A, every asymptotically reliable sequence of sensing operators satisfies

\begin{equation}
	\label{37}
	\underset{m \rightarrow \infty}{p - liminf}\ M_{m}\  \geq \ M_{H}(\Xi \mid Y) = 2^{\mathcal{H}(\Xi \mid Y)},
\end{equation}

and the typical-set sensor attains this lower bound. 
The minimum achievable posterior entropy volume is
\begin{equation}
	\label{38}
	\begin{aligned}
		M_H^{\star}
		&= 2^{\mathcal{H}(\Xi)}\,2^{-I(\Xi;Y)} \\
		&= N_H(V\mid Y)\,
		V_H(X_V\mid V,Y).
	\end{aligned}
\end{equation}

\noindent\textit{Proof.}
Lemma 5 directly gives the probability lower limit of the empirical
entropy volume for any asymptotically reliable sensing operator.
Monotonicity and continuity of the exponential function yield (37).
Conversely, Lemma 3(ii) shows that the empirical entropy volume of the
typical-set sensor converges in probability to
$2^{\mathcal{H}(\Xi\mid Y)}$, so the lower bound is achievable.
Finally, using
$\mathcal{H}(\Xi\mid Y)
=\mathcal{H}(\Xi)-I(\Xi;Y)$
together with the mixed entropy-volume decomposition in Section~III
gives (38).
\hfill$\square$

\subsection{Information Limit: Logarithmic Form of Entropy-Volume Contraction}

\noindent\textbf{Theorem 2 (Information Limit of Joint Detection and
	Parameter Estimation).}
Under Assumption A, define
\[
\begin{aligned}
	J^{\star}\triangleq
	\sup\bigl\{J:\;&
	\text{there exists an asymptotically reliable sequence}\\
	&
	\text{of sensing operators such that }
	\operatorname*{p\text{-}liminf}_{m\to\infty}J_m\geq J
	\bigr\}.
\end{aligned}
\]
Then
\begin{equation}
	\label{39}
	J^{\star}=I(\Xi;Y).
\end{equation}

\noindent\textit{Proof.}
By definition, empirical sensing information equals the prior per-state
log entropy volume minus the empirical posterior log volume.
Theorem~1 gives the minimum achievable posterior entropy volume
$2^{\mathcal{H}(\Xi\mid Y)}$ for every asymptotically reliable sensing
region. Taking logarithms and using
$\mathcal{H}(\Xi)-\mathcal{H}(\Xi\mid Y)=I(\Xi;Y)$
gives
$J^{\star}\leq I(\Xi;Y)$.
The typical-set sensor attains the entropy-volume lower bound in
Theorem~1 and therefore also attains
$J^{\star}=I(\Xi;Y)$.
Thus the information limit is not a separate theorem independent of
entropy volume; it is the logarithmic representation of the same
posterior-volume limit.
\hfill$\square$

The data-processing inequality can separately constrain the information
in point or compressed outputs (see Section~VI), but it cannot replace
the region achievability and region converse supplied by Lemmas~3 and~5.

Theorems~1 and~2 are in one-to-one correspondence through
$M_H=2^{\mathcal{H}}$. They are two representations of the same limit
in the physical-measure domain and the logarithmic information domain,
rather than two independent limits. The optimal posterior entropy
volume is determined by $2^{\mathcal{H}(\Xi\mid Y)}$, while
$I(\Xi;Y)$ is the dimensionless logarithmic contraction rate from the
prior entropy volume to that limit. Hence each additional bit of sensing
information halves the optimal posterior entropy volume. The two
representations have complementary roles: volume carries physical units,
can be compared directly with engineering scales such as resolution
cells, and remains well defined when the number of active targets
varies; information is independent of coordinates and units, directly
comparable with communication information, and additive through the
chain rule.

\noindent\textbf{Note 2.}
Here ``minimum achievable'' always means the first-order asymptotic
scale of the per-state reference measure under an $m$-fold memoryless
extension, not exact samplewise minimization over all credible regions
at an arbitrary finite $m$. The typical-set sensor provides both a
high-probability region attaining this first-order scale and a
posterior-sampled representative sequence having the same distribution
as the true state sequence.

\subsection{Entropy Bound for Nonvanishing Failure Probability}

The posterior entropy-volume converse above assumes asymptotic
reliability, $P_f^{(m)}\to 0$. When the failure probability does not
vanish, a high-probability region no longer obeys the same direct
conclusion. In that case, the maximum-entropy principle of Section~III
can be combined with an error-event decomposition to obtain an
entropy-domain bound with a failure-probability penalty. This result
complements the main posterior-volume converse for finite failure
probability and is not a premise of Theorems~1 or~2.

\noindent\textbf{Assumption B.}
$\mu(\mathcal{S})<\infty$.

Assumption B holds automatically for a bounded location domain and a
finite number of branches. Let
\[
\alpha_m
\triangleq
\Pr\!\left\{
\Xi^m\notin A_{\varepsilon}^{(m)}(\Xi)
\right\};
\]
by Lemma~1(i), $\alpha_m\to0$. Introduce the ternary indicator
\begin{equation}
	E=
	\begin{cases}
		0, &
		\Xi^m\in\mathcal{R}_m,\quad
		\Xi^m\in A_{\varepsilon}^{(m)}(\Xi), \\[2pt]
		1, &
		\Xi^m\notin\mathcal{R}_m,\quad
		\Xi^m\in A_{\varepsilon}^{(m)}(\Xi), \\[2pt]
		2, &
		\Xi^m\notin A_{\varepsilon}^{(m)}(\Xi).
	\end{cases}
	\label{eq:ternary_indicator}
\end{equation}

\noindent\textbf{Proposition 1 (Mixed Generalized Fano Inequality).}
Under Assumptions A and B, use the error-event decomposition of Fano's
inequality~\cite{ref32}, but control the maximum entropy on the success
branch by the mixed measure of the sensing region rather than by a
number of candidate messages. Then
\begin{equation}
	\begin{aligned}
		&\mathcal{H}(\Xi\mid Y)\\
		&\leq{}
		\frac{\log_2 3}{m}
		+\Pr(E=0)\frac{1}{m}
		\mathbb{E}\!\left[
		\log_2\mu^m\!\left(\mathcal{R}_m(Y^m)\right)
		\mid E=0
		\right]
		\\
		&+\Pr(E=1)
		\left[\mathcal{H}(\Xi)+\varepsilon\right]
		+\alpha_m\log_2\mu(\mathcal{S}).
	\end{aligned}
	\label{eq:mixed_generalized_fano}
\end{equation}

\noindent\textit{Proof.} By the entropy chain rule, \(\mathcal{H}\left( \Xi^{m} \mid Y^{m} \right) \leq H(E)\mathcal{+ H}\left( \Xi^{m} \mid Y^{m},E \right)\), where \(H(E) \leq \log_{2}3\). On branch \(E = 0\), conditional on \(Y^{m}\), \(R_{m}\left( Y^{m} \right)\) is a deterministic set containing the true state. The maximum-entropy principle relative to \(\mu^{m}\) gives \(\mathcal{H}\left( \Xi^{m} \mid Y^{m} = y^{m},E = 0 \right) \leq \log_{2}\mu^{m}\left( R_{m}\left( y^{m} \right) \right)\); take the conditional expectation over \(y^{m}\). On branch \(E = 1\), the state lies in the prior typical set, so (17) gives \(\mathcal{H}\left( \Xi^{m} \mid Y^{m},E = 1 \right) \leq m\left\lbrack \mathcal{H}(\Xi) + \varepsilon \right\rbrack\). On branch \(E = 2\), the state can only be bounded by the full state space; Assumption B gives \(\mathcal{H}\left( \Xi^{m} \mid Y^{m},E = 2 \right) \leq m\log_{2}\mu(S)\), and \(\Pr(E = 2) = \alpha_{m}\). Divide by \(m\) and use \(\mathcal{H}\left( \Xi^{m} \mid Y^{m} \right) = m\mathcal{H}(\Xi \mid Y)\) to obtain (41). \hfill$\square$

The failure event must be split further according to typicality: conditioning on \ensuremath{\Xi}\^{}m\ensuremath{\notin}R\_m does not guarantee that \ensuremath{\Xi}\^{}m remains in the prior typical set. Hence only branch E=1 can be bounded by the prior typical-set measure; branch E=2 must be bounded by the entire state space, and its coefficient \ensuremath{\alpha}\_m\ensuremath{\rightarrow}0 makes the term vanish asymptotically. When P\_f\^{}\{(m)\}\ensuremath{\rightarrow}0 and \ensuremath{\alpha}\_m\ensuremath{\rightarrow}0, (41) is of the same first order as Lemma 5. When P\_f\^{}\{(m)\} remains finite, the term Pr(E=1){[}\ensuremath{\mathcal{H}}(\ensuremath{\Xi})+\ensuremath{\varepsilon}{]} quantifies the residual-entropy penalty.

\section{Posterior-Preserving Cascades and Information Loss from Intermediate Compression}

\subsection{Posterior Factorization and Lossless Cascades}

For the mixed state \(\Xi = \left( V,X_{V} \right)\), the joint posterior factorizes exactly as

\begin{equation}
	\label{42}
	p\left( v,x_{v} \mid y \right) = p(v \mid y)\, p\left( x_{v} \mid v,y \right).
\end{equation}

A two-stage architecture does not by itself cause information loss. Traditional joint detection--estimation studies optimal rules for one-step or two-step decision systems \cite{ref26}; our concern is whether the interstage representation preserves the complete posterior information. Classical results on sufficiency and comparison of experiments in statistical decision theory show that a statistic preserving all decision-relevant information about the unknown state can be regarded as information-equivalent \cite{ref33}.

\noindent\textbf{Theorem 3 (Losslessness of a Posterior-Preserving Cascade).} Let the intermediate cascade representation \(Z_{pp}\) be generated from \(Y\) and satisfy \(p(\Xi \mid Y) = p\left( \Xi \mid Z_{pp} \right)\) a.s. Then

\begin{enumerate}
\def\labelenumi{(\roman{enumi})}
\item
\(I\left( \Xi;Z_{pp} \right) = I(\Xi;Y)\);
\item
For an \(m\)-fold extension, let the detection stage output \(D_{m} = A_{\varepsilon}^{(m)}\left( V \mid Y^{m} \right)\), and for every \(v^{m}\) in \(D_{m}\) assign the conditional estimation region \(E_{m}\left( v^{m} \right) = A_{\varepsilon}^{(m)}\left( X_{V} \mid Y^{m},v^{m} \right)\). Then the cascaded sensing region
\end{enumerate}

\begin{equation}
	\label{43}
	R_{m}^{cas} = \bigsqcup_{v^{m} \in D_{m}}\{ v^{m}\} \times E_{m}\left( v^{m} \right)
\end{equation}

\begin{itemize}
\item
satisfies \(\mu^{m}\left( R_{m}^{cas} \right) = 2^{m\left\lbrack \mathcal{H}(\Xi \mid Y) + o(1) \right\rbrack}\), and therefore
\end{itemize}

\begin{equation}
	\label{44}
	J_{m}^{cas}\  \rightarrow \ I(V;Y) + I\left( X_{V};Y \mid V \right) = I(\Xi;Y),
\end{equation}

\begin{itemize}
\item
Thus the cascade attains the same limits as Theorems 1 and 2.
\end{itemize}

\noindent\textit{Proof.} (i) Because \(Z_{pp}\) is generated from \(Y\), \(\Xi \rightarrow Y \rightarrow Z_{pp}\) forms a Markov chain. By assumption, \(Z_{pp}\) is sufficient to recover \(p(\Xi \mid Y)\), so \(I\left( \Xi;Y \mid Z_{pp} \right) = 0\). Substitution into the chain rule \(I(\Xi;Y) = I\left( \Xi;Z_{pp} \right) + I\left( \Xi;Y \mid Z_{pp} \right)\) proves the claim.

(ii) The branches in (43) are disjoint, so the mixed measure is the sum of the branch measures. For typical \(v^{m}\), the decision-region cardinality has exponential order \(2^{mH(V \mid Y)}\), while the volume of each conditional estimation region has exponential order determined by \(h\left( X_{V} \mid Y,V \right)\). The total continuous dimension \(d\sum_{i}^{}\left| v_{i} \right|\) varies with \(v^{m}\), but by the law of large numbers it concentrates around \(md\,\mathbb{E}|V|\). The exponential orders can therefore be combined, giving

\begin{equation}
	\label{45}
	\mu^{m}\left( R_{m}^{cas} \right) = 2^{m\left\lbrack H(V \mid Y) + h\left( X_{V} \mid Y,V \right) + o(1) \right\rbrack} = 2^{m\left\lbrack \mathcal{H}(\Xi \mid Y) + o(1) \right\rbrack}.
\end{equation}

Taking logarithms, dividing by \(m\), and using Definition 5 gives (44). \hfill$\square$

In Theorem 3(ii), the second stage assigns a conditional estimation region to every \(v^{m}\) in the decision region, rather than only to a selected \({\widehat{V}}^{m}\). Consequently, (45) contains the conditional entropy \(h\left( X_{V} \mid Y,V \right)\) rather than \(h\left( X_{V} \mid Y,\widehat{V} \right)\), and the chain decomposition holds exactly. This is the distinction between a posterior-preserving cascade and a detect-then-estimate procedure based on an early hard decision.

\noindent\textit{Note 3 (Combining Branches in (45)).}
Different \(v^m\in D_m\) contain different numbers of active targets and therefore have different continuous dimensions; their \(\mu^m\) measures cannot be compared term by term as if they shared a common dimension. Equation (45) should instead be understood as the first-order exponential rate of the sum of branch measures, verified separately by upper and lower bounds. For the upper bound, the cardinality of \(D_m\) is at most \(2^{m[H(V\mid Y)+\varepsilon]}\) to exponential order, while the conditional typical-set measure bound gives at most \(2^{m[h(X_V\mid Y,V)+\varepsilon]}\) for each \(E_m(v^m)\); their product yields the upper bound in (45), irrespective of whether a particular \(v^m\) has a typical activity count. For the lower bound, it suffices to retain only conditionally typical \(v^m\) whose activity counts lie near their concentration value; both their number and the measure of each associated branch obey the corresponding exponential lower bounds. Branches away from the concentration value therefore neither invalidate the upper bound nor need to enter the lower bound, and the single exponent in (45) follows.

\subsection{Information Loss under Arbitrary Intermediate Compression}

\noindent\textbf{Proposition 2 (Information-Loss Identity).} Let the intermediate cascade representation be \(Z = g(Y,U)\), where \(U\) is independent of \((\Xi,Y)\), and suppose the downstream detector and estimator can access only \(Z\). Then

\begin{equation}
	\label{46}
	\Delta I_{cas} \triangleq I(\Xi;Y) - I(\Xi;Z) = I(\Xi;Y \mid Z)\  \geq \ 0,
\end{equation}

and

\begin{equation}
	\label{47}
	\Delta I_{cas} = 0\  \Leftrightarrow \ I(\Xi;Y \mid Z) = 0\  \Leftrightarrow \ p(\Xi \mid Y) = p(\Xi \mid Z)\ \text{a.s.}
\end{equation}

\noindent\textit{Proof.} Since \(Z = g(Y,U)\) and \(U\) are independent, \(I(\Xi;Z \mid Y) = 0\), and hence \(I(\Xi;Y,Z) = I(\Xi;Y) + I(\Xi;Z \mid Y) = I(\Xi;Y)\). Expanding the same joint mutual information in the other order gives \(I(\Xi;Y,Z) = I(\Xi;Z) + I(\Xi;Y \mid Z)\). Equating the two expressions yields (46), and nonnegativity follows from nonnegativity of mutual information. In (47), the first equivalence follows from the definition of \(\Delta I_{cas}\); the second is the equivalence between zero conditional mutual information and conditional independence, which in the present controlled model is \(p(\Xi \mid Y) = p(\Xi \mid Z)\) a.s. \hfill$\square$

Proposition 2 refines the data-processing inequality \cite{ref2}. The latter gives \(I(\Xi;Z) \leq I(\Xi;Y)\), whereas Proposition 2 expresses the gap in that inequality as an exact residual information term. Whether a cascade is lossless is determined not by its two-stage computational structure, but by whether the interstage representation is sufficient for the complete mixed state.

\subsection{Hard-Decision Cascades}

If the first stage compresses the posterior into a hard decision \(\widehat{V}\), or passes only \(\left( \widehat{V},T(Y) \right)\) to the second stage, then unless that output is itself sufficient for \(\Xi\), one generally has \(I\left( \Xi;\widehat{V},T(Y) \right) < I(\Xi;Y)\). In particular, transmitting only \(\widehat{V}\) is usually insufficient to preserve the observation information needed for the continuous parameters \(X_{V}\). The architecture-level performance gap can therefore be reduced to posterior sufficiency: if the interstage representation is sufficient for \(\Xi\), the cascade is lossless; hard decisions or coarse quantization are lossy because they are generally not sufficient statistics. This conclusion does not invalidate traditional two-step joint detection--estimation \cite{ref26}; rather, it identifies the source of its possible performance gap.

\section{Detection and Estimation as Two Boundary Corollaries}

\noindent\textbf{Corollary 1 (Pure Parameter Estimation).} If \(P\left( V = v_{0} \right) = 1\), then Theorems 1 and 2 reduce to

\begin{equation}
	\label{48}
	J^{*} = I(X;Y),\quad\quad V_{H}^{\min}(X) = V_{H}(X \mid Y) = V_{H}(X)\, 2^{- I(X;Y)}.
\end{equation}

\noindent\textit{Proof.} In this case \(H(V) = H(V \mid Y) = I(V;Y) = 0\), so the branched state contains only the deterministic presence pattern \(v_{0}\). Mixed entropy, the mixed typical set, and the mixed measure reduce respectively to ordinary differential entropy, a continuous typical set, and Lebesgue measure. Lemmas 1--5 and Theorems 1 and 2 apply verbatim. \hfill$\square$

The information limit for parameter estimation is therefore the continuous boundary case of the joint theorem. Unlike classical parameter-estimation theory \cite{ref25}, the present formulation first establishes information and volume limits for posterior regions and treats the risk of a particular estimator as a subsequent decision-layer quantity.

\noindent\textbf{Corollary 2 (Pure Target Detection).} If the state contains no continuous uncertainty, so that \(\Xi = V\), then Theorems 1 and 2 reduce to

\begin{equation}
	\label{49}
	J^{*} = I(V;Y), N_{H}^{\min}(V) = N_{H}(V \mid Y) = N_{H}(V)\, 2^{- I(V;Y)}.
\end{equation}

\noindent\textit{Proof.} Here \(X_{V}\) is an empty state, \(h\left( X_{V} \mid V,Y \right) = 0\), the branched reference measure reduces to counting measure, and \(M_{H}(\Xi \mid Y) = 2^{H(V \mid Y)} = N_{H}(V \mid Y)\). Theorems 1 and 2 apply verbatim. \hfill$\square$

The information limit for target detection is therefore the discrete boundary case of the same joint theorem. Classical hypothesis-testing performance can still be evaluated by error probability and the Neyman--Pearson criterion \cite{ref25}; the quantity characterized here is the information contraction of a random physical source under a fixed prior after observation. Neither corollary requires a new achievability or converse proof; both follow directly by degeneration of the state space and reference measure.

\section{Numerical Results}

This section does not treat analytic identities themselves as objects to be ``validated by simulation.'' Instead, the numerical results illustrate the physical meaning of posterior entropy volume as a scale carrying physical units and its consequences for receiver architecture. The detection--estimation composition of joint information determines the same total contraction rate; posterior entropy volume contracts by the exponential factor prescribed by joint mutual information; and an insufficient cascade interface incurs quantifiable information and volume loss.

\subsection{Numerical Model and Implementation}

We use the single-target presence--range model

\begin{equation}
	\label{50}
	y = V\, S\, u(X) + w,
\end{equation}

where \(V \in \{ 0,1\}\) with \(P(V = 1) = 1/2\), the continuous location parameter \(X\) is uniform on the normalized interval \(\lbrack 0,32)\), and \(w\sim\mathcal{CN}(0,I)\). The main experiment uses constant-modulus scattering \(S = \sqrt{\gamma}\, e^{j\Phi}\) with \(\Phi \sim U\lbrack 0,2\pi)\), where \(\gamma\) is the receive SNR. This benchmark follows the single-target ranging model used in the authors' previous work on range information and sensing information \cite{ref13}, \cite{ref15}, allowing mixed-state information decomposition, the posterior-volume law, and the cascade-loss identity to be examined on the same physical baseline.

The sensing mode is a unit-energy bandlimited ranging waveform. Noise is generated first in the observation domain and then projected onto the sensing-mode bank, preserving the waveform-correlation-induced dependence among matched-filter samples. The continuous posterior is evaluated numerically on \(G = 1024\) grid points, and every differential-entropy calculation includes the grid-spacing correction \(\log_{2}\Delta x\). Each SNR point uses \(2 \times 10^{4}\) Monte Carlo realizations, with the 0--16 dB transition region sampled every 2 dB. Because the position coordinate is normalized by the resolution cell, the prior location uncertainty is \(h(X)=5\) bits, corresponding to an entropy volume \(V_H(X\mid V=1)=32\) resolution cells. The entropy volumes below can therefore be read directly in units of resolution cells.

For \(V = 0\), \(p_{0}(y) = \pi^{- N}\exp\left( - \lVert y\rVert^{2} \right)\). For \(V = 1\) and a given \(X = x\), integration over the uniform random phase gives the constant-modulus likelihood

\begin{equation}
	\label{51}
	p_{1}(y \mid x)\  \propto \ \exp\left\lbrack - \left( \lVert y\rVert^{2} + \gamma \right) \right\rbrack\, I_{0}\left( 2\sqrt{\gamma}\,\left| u(x)^{H}y \right| \right),
\end{equation}

where \(I_{0}( \cdot )\) is the modified Bessel function of the first kind. The numerical implementation uses a scaled Bessel function and log-sum-exp to avoid overflow at high SNR. Under the uniform prior, \(p(y \mid V = 1)\) is obtained by quadrature over the \(x\) grid, from which the presence posterior \(q(y) = P(V = 1 \mid y)\) and location posterior \(p(x \mid y,V = 1) \propto p_{1}(y \mid x)p_{X}(x)\) are computed. Detection information is evaluated from \(I_D=H(V)-\mathbb{E}[H_b(q(Y))]\), and conditional estimation information from \(I_E=\pi_1\left[h(X)-\mathbb{E}\{h(X\mid Y,V=1)\}\right]\). To independently check the chain decomposition, \(I_{J}\) is also computed directly from \(\mathbb{E}\left\lbrack \log_{2}p(Y \mid \Xi)/p(Y) \right\rbrack\).

\subsection{Decomposition of Joint Sensing Information}

Fig.~1 shows \(I_D\), \(I_E\), and \(I_J\) versus SNR. They are computed independently from the binary posterior entropy, the continuous posterior differential entropy, and the joint likelihood, respectively. Over the full SNR range, the maximum Monte Carlo discrepancy between directly computed \(I_J\) and \(I_D+I_E\) is approximately \(4.2\times10^{-2}\) bits, about 1\% of \(I_J\), providing an independent numerical check of (12). At low SNR all three quantities approach zero. As SNR increases, the detection information approaches \(H(V)=1\) bit while the continuous-information term continues to grow, so at high SNR nearly all additional sensing information is used to compress location uncertainty. At 12 dB, \(I_D=0.882\) bits, \(I_E=3.119\) bits, and \(I_J=4.001\) bits.

\begin{figure}[t]
\centering
\includegraphics[width=0.6\linewidth]{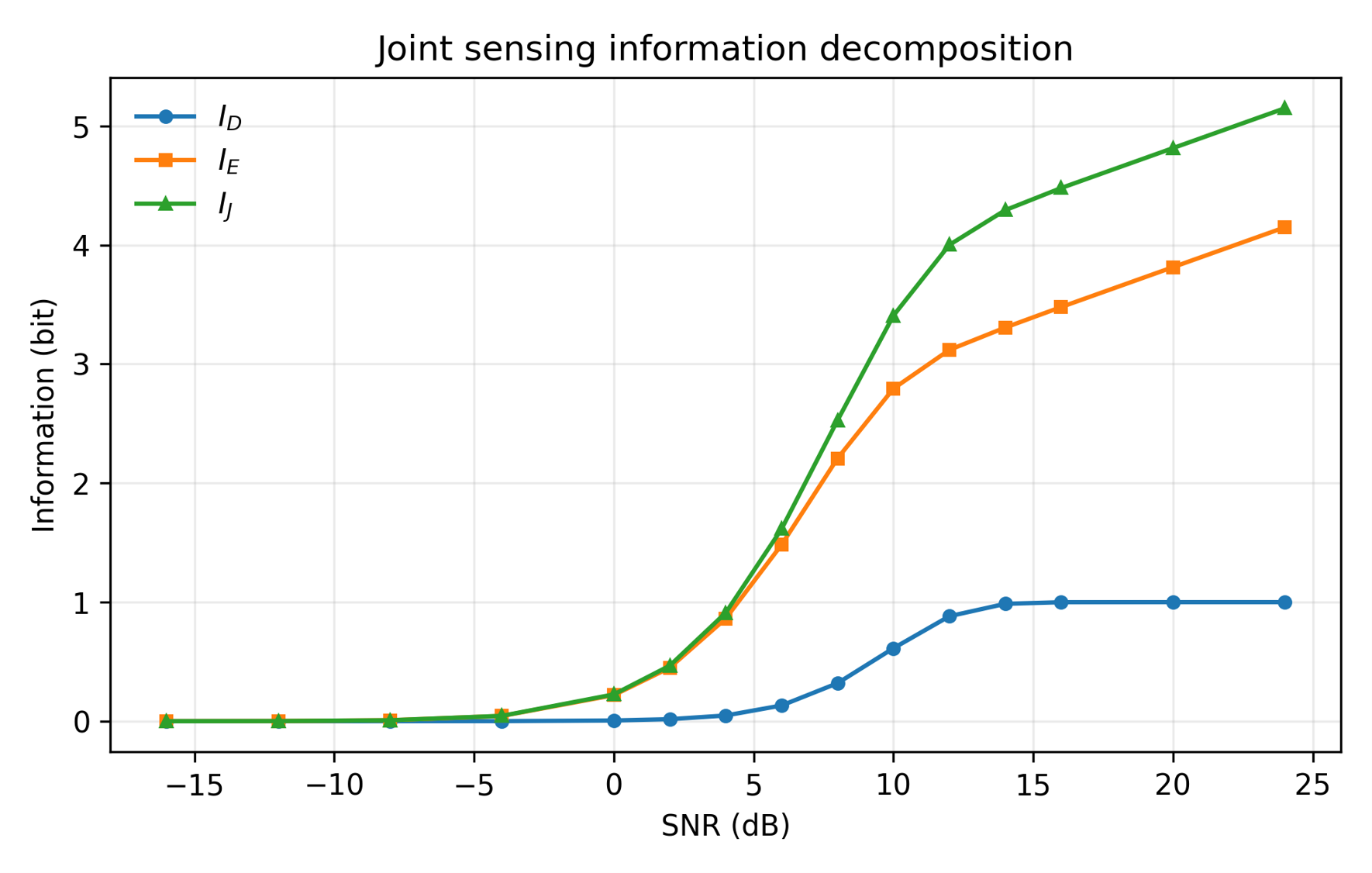}
\caption{Detection information, conditional estimation information, and joint sensing information versus SNR.}
\label{fig:1}
\end{figure}

\subsection{Posterior Entropy-Volume Contraction}

Fig.~2 presents the same process in the physical-uncertainty domain. From (10) and (13),
\begin{equation}
	\label{51}
	\frac{M_H(\Xi\mid Y)}{M_H(\Xi)}
	=
	\frac{N_H(V\mid Y)}{N_H(V)}
	\cdot
	\frac{V_H(X\mid Y,V)}{V_H(X\mid V)}
	=
	2^{-I_J}.
\end{equation}

Equation (52) is an identity when both sides are computed from the same underlying quantities. In Fig.~2, the left-hand side is obtained directly from posterior entropies, while \(I_J\) on the right-hand side is computed independently as described in Section VIII-A. The two curves coincide to floating-point precision, serving as a consistency check; the independent numerical checks are the chain decomposition in Fig.~1 and the cascade loss in Fig.~4. The role of Fig.~2 is to place addition in the information domain alongside multiplication in the physical domain: detection and estimation information add in bits, while the corresponding uncertainty-contraction factors multiply in the entropy-volume domain. Fig.~3 gives the engineering interpretation of the same relation by calibrating posterior entropy volume in resolution cells, so the two figures serve different purposes.

\begin{figure}[t]
\centering
\includegraphics[width=0.6\linewidth]{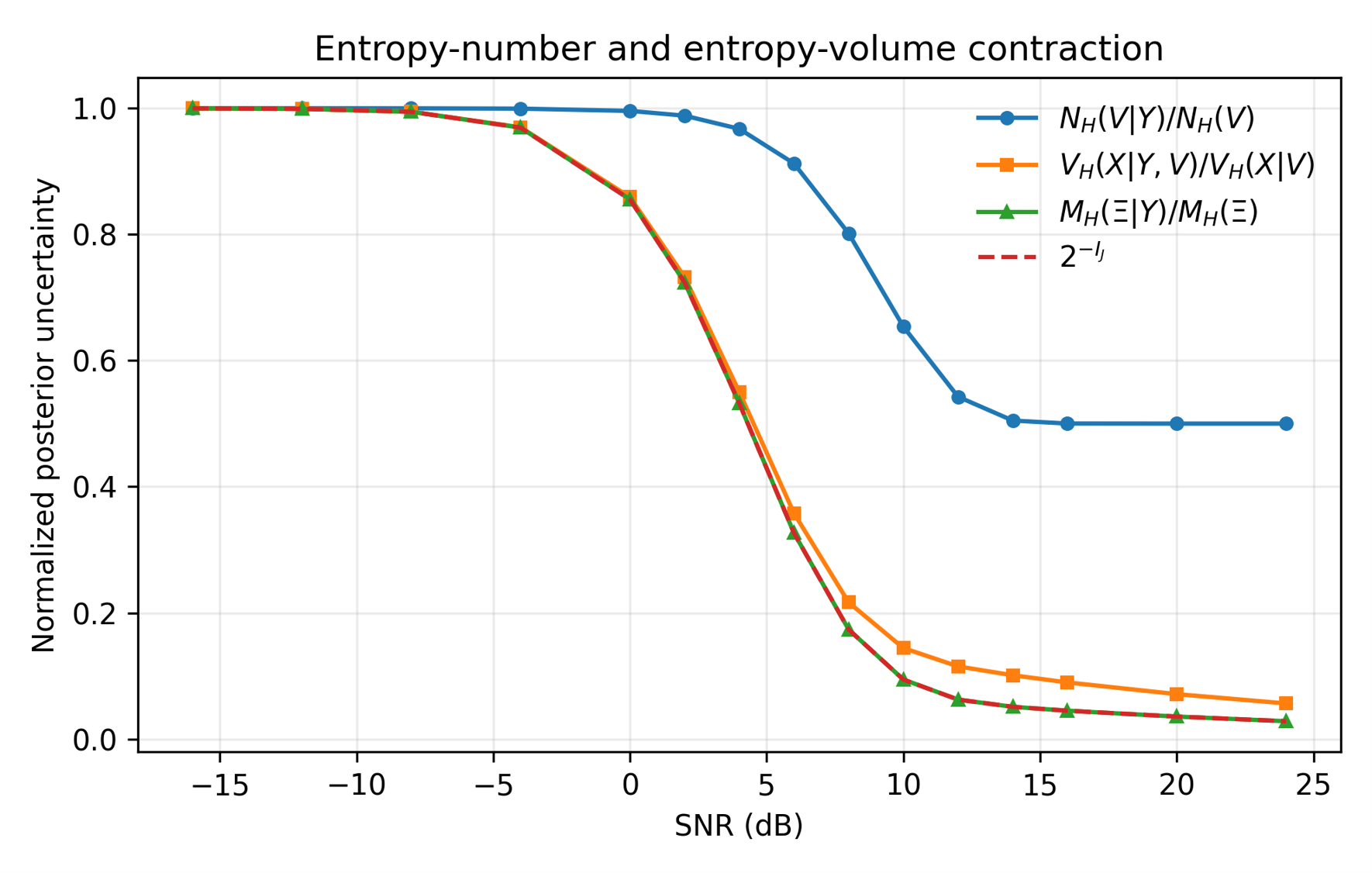}
\caption{Normalized contraction of entropy number, conditional entropy volume, and mixed posterior entropy volume.}
\label{fig:2}
\end{figure}

\subsection{Physical Meaning of One Sensing Bit}

Fig.~3 plots normalized posterior entropy volume against independently computed \(I_J\). The numerical points lie on the theoretical curve \(M_H(\Xi\mid Y)/M_H(\Xi)=2^{-I_J}\): 1, 2, 3, and 4 sensing bits reduce the posterior entropy volume to \(1/2\), \(1/4\), \(1/8\), and \(1/16\) of its prior value, respectively. The same result is even more intuitive when read in resolution cells. In this model, the prior location uncertainty is 32 resolution cells. At 12 dB, the posterior entropy volume conditioned on target presence is
\[
V_H(X\mid Y,V=1)
=
2^{h(X)-I_E/\pi_1}
\approx 0.42
\]
resolution cells, so the location posterior has contracted to less than one resolution cell. At the same operating point, the entropy number is \(N_H(V\mid Y)=2^{H(V\mid Y)}\approx 1.09\), meaning that the effective number of candidates for target absence/presence is already close to one. This is a direct advantage of entropy volume over mutual information: it has the same physical scale as radar resolution cells and unambiguous windows and can be compared with them without conversion, whereas ``3.119 bits'' must first be translated into a physical scale before it can be interpreted as an engineering quantity.

\begin{figure}[t]
\centering
\includegraphics[width=0.6\linewidth]{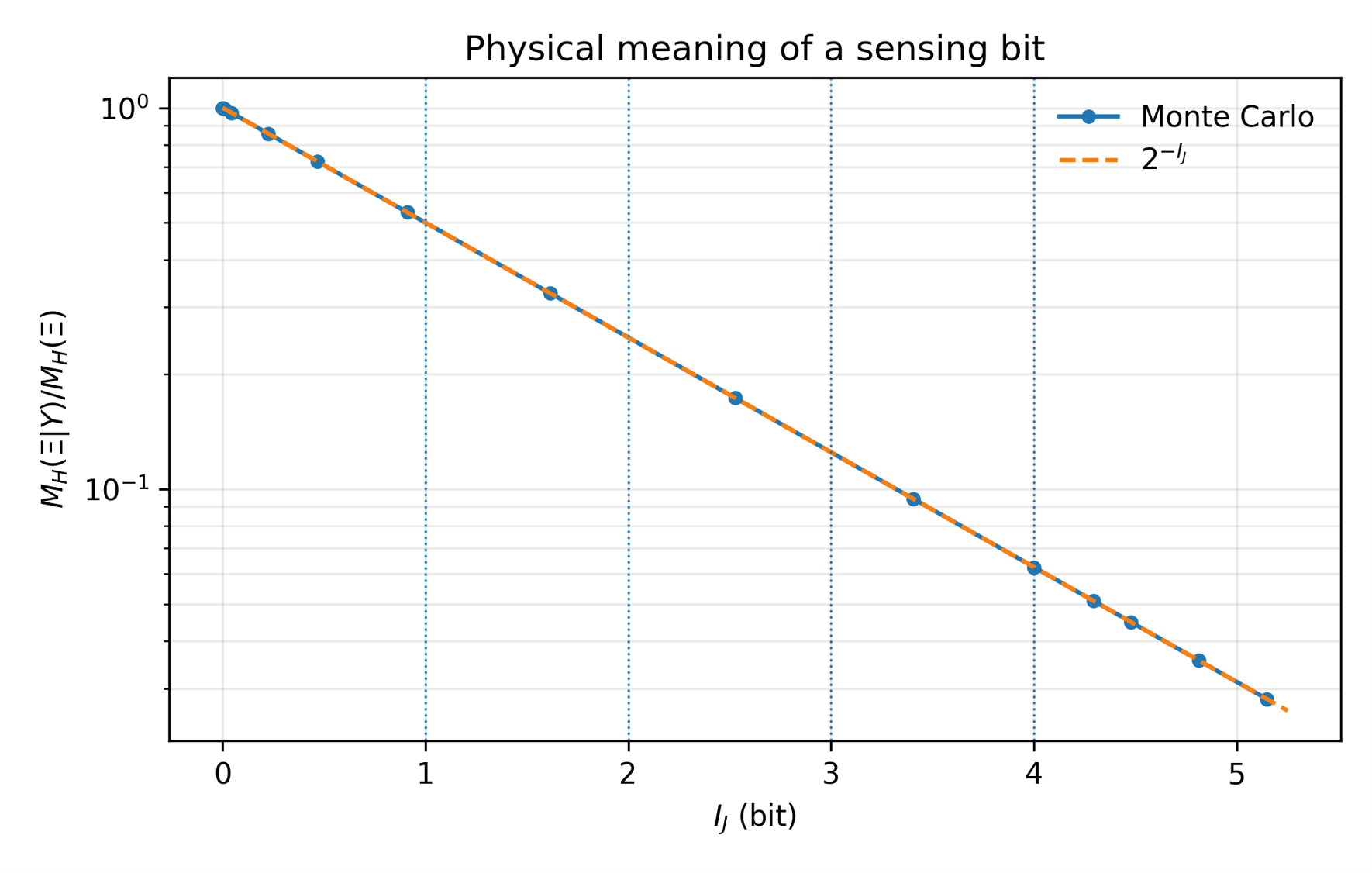}
\caption{Relationship between joint sensing information and posterior entropy-volume contraction.}
\label{fig:3}
\end{figure}

\subsection{Interface Information Loss in Cascaded Receivers}

Fig.~4 corresponds to Section VI. We compare three interfaces: the ideal posterior-preserving interface, an interface passing only the hard decision \(Z=\widehat V\), and interfaces also passing a quantized location output \(Z=(\widehat V,\widehat X_Q)\) with \(Q=16,64,256\). The theoretical loss is given by (46). Numerically, \(I(\Xi;Z)\) is evaluated as \(I(\Xi;Z)=\mathcal{H}(\Xi)-\mathcal{H}(\Xi\mid Z)\), where \(Z\) is discrete and its distribution and conditional posterior are estimated from the Monte Carlo realizations. At 12 dB, \(I_J\approx 4.001\) bits; the corresponding losses are approximately 3.20 bits for \(\widehat V\) only, 1.31 bits for \(Q=16\), 0.52 bits for \(Q=64\), and 0.29 bits for \(Q=256\). As the interface resolution increases, the loss decreases markedly and approaches the zero-loss limit of the posterior-preserving interface.

These results show that the performance gap of a hard-decision cascade comes from compressing the posterior into a low-dimensional point output at the first stage, not from the computational order of detecting first and estimating second. By the data-processing inequality \cite{ref2}, subsequent processing cannot recover information about \(\Xi\) that has been discarded by an insufficient interface, consistent with the classical conclusions on sufficiency and comparison of experiments \cite{ref33}.

\begin{figure}[t]
\centering
\includegraphics[width=0.6\linewidth]{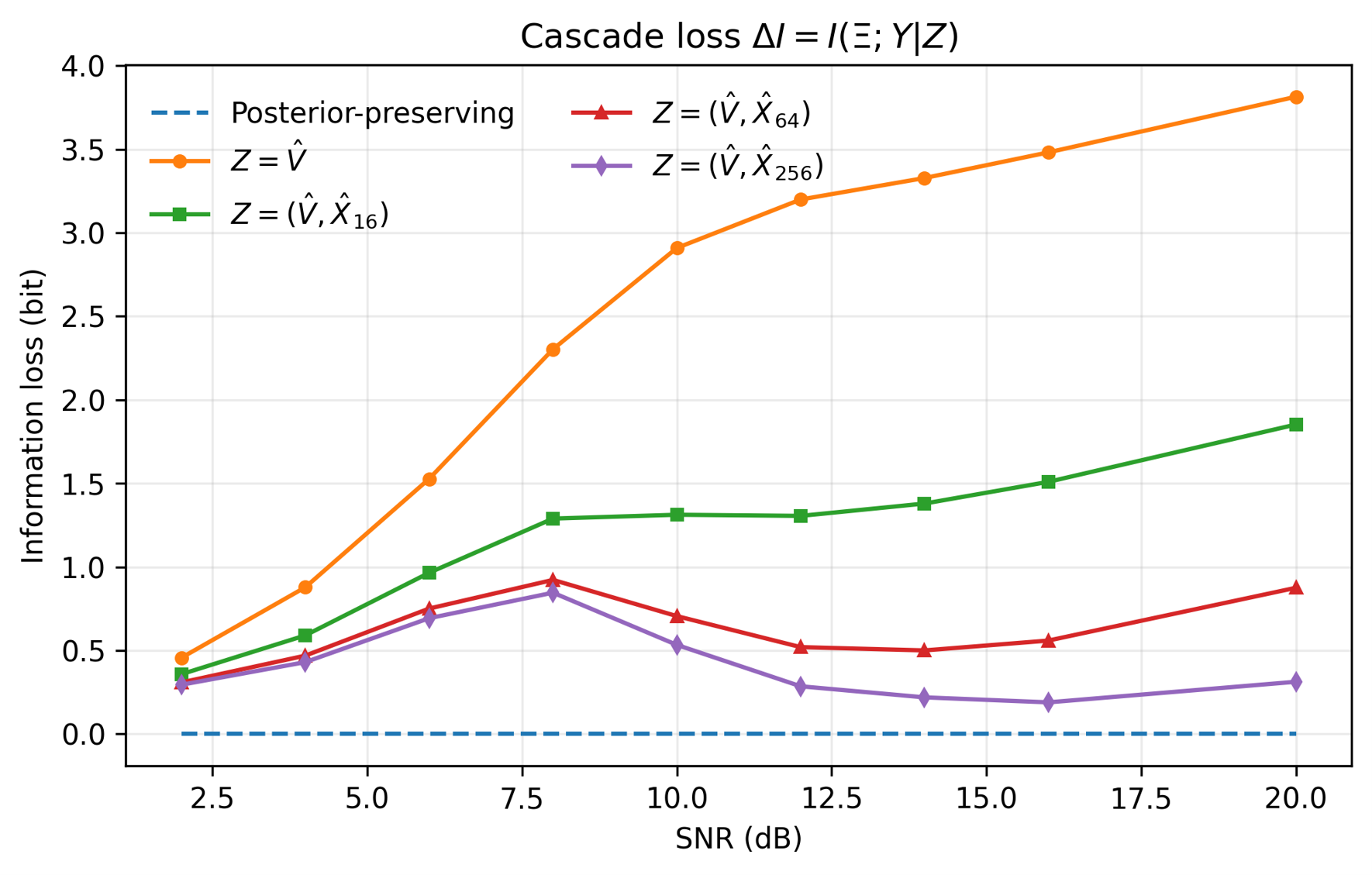}
\caption{Information loss \(\Delta I_{\mathrm{cas}}\) under different cascade-interface resolutions. The theoretical loss of a posterior-preserving cascade is zero.}
\label{fig:4}
\end{figure}

\subsection{Numerical Convergence}

At 4, 8, 12, and 16 dB, grid convergence of \(I_{J}\) is checked using \(G = 512,1024,2048\). The three results agree closely over the main operating range; at 12 and 16 dB, the difference between \(G = 1024\) and \(G = 2048\) is on the order of \(10^{-3}\). All differential-entropy calculations include the grid-spacing correction, so the entropy volume does not undergo an artificial shift when the grid density changes. These results correspond respectively to (12), (13), and (46).

\section{Discussion: Measures, Dimensions, and Scope}

The fundamental objects in this paper are the entropy volume \(V_{H} = 2^{h}\) and mixed entropy volume \(M_{H} = 2^{\mathcal{H}}\) relative to a prescribed reference measure; the reference-measure dependence of entropy on general alphabets is discussed in \cite{ref28}. Gaussian calibration is used only within fixed-dimensional continuous branches to connect with covariance, MMSE and Cramér--Rao-type quantities \cite{ref27}, and with the authors' previous joint entropy-error results \cite{ref18}, \cite{ref19}. The unified theorem for variable-dimensional mixed states does not require imposing a common Euclidean dimension on all branches. Entropy volume depends on the reference measure associated with the physical coordinates, whereas mutual information is the difference between prior and posterior log-measures on the same state space. We therefore continue to use ordinary Shannon bits; a sensing bit refers to logarithmic compression of posterior uncertainty measure. The reference-measure dependence of entropy volume does not make it secondary to mutual information: the reference measure is specified by the physical measurement system and is part of the model. Volume and information are two readings of the same limit---the former is directly comparable with engineering scales and applies to variable-dimensional mixed states, while the latter is coordinate invariant and additive.

The achievability result is asymptotic under an \(m\)-fold independent extension, asymptotic equipartition, and a sensing-region formulation \cite{ref2}, \cite{ref29}. Posterior sampling is a construction used to prove achievability; it does not imply that a practical finite-dimensional system must produce a randomized output, nor that posterior sampling is superior to MAP or MMSE estimation under a particular Bayesian risk \cite{ref25}. Finite blocklength, non-i.i.d. extensions, model mismatch, and computational complexity are left for future work.

The finite-order multi-target model does not cover an arbitrary random number of targets, unlabeled states, permutation equivalence classes, or data association. The theorem has potential extensions to more general mixed measures, but the random-finite-set case requires separate treatment of its reference measure, typical sets, and association uncertainty.

\section{Conclusion}

This paper studied the posterior entropy-volume limit of joint target detection and parameter estimation for mixed discrete--continuous states. The uniform maximum-entropy principle on a support of fixed measure first shows that exponential entropy gives the minimum effective support-measure scale compatible with a prescribed entropy. The mixed AEP then shows that the conditional posterior has the corresponding exponential equipartition on a high-probability conditional typical set, and the posterior probability--volume converse lifts this geometric relation into an achievability--converse theorem for asymptotically reliable sensing. Consequently, the minimum achievable posterior mixed entropy volume is determined by the exponential scale of the conditional mixed entropy, while the joint mutual information \(I(\Xi;Y)\) is the logarithmic contraction rate from the prior entropy volume to this limit. Detection information and conditional estimation information add in the bit domain, whereas entropy number and conditional entropy volume multiply in the physical-measure domain; one sensing bit halves the optimal posterior effective measure. A posterior-preserving cascade attains the same limit, while the loss due to arbitrary intermediate compression is exactly \(I(\Xi;Y\mid Z)\). Joint detection and parameter estimation are therefore unified as posterior-volume contraction of a single mixed physical state. In one sentence: the fundamental limit of sensing is the achievable posterior volume under the physical reference measure, mutual information is the dimensionless logarithmic reading of that volume contraction, and detection and estimation multiply in the volume domain while adding in the bit domain.

{\appendices


\section{Biography Section}

\begin{IEEEbiographynophoto}{Dazhuan Xu}
	is a Professor jointly affiliated with Nanjing University of Aeronautics and Astronautics and Purple Mountain Laboratories, Nanjing, China. His research interests include information theory for sensing, spatial information theory, broadband wireless communications, and statistical signal processing. His recent research has focused on the information-theoretic foundations of sensing, including fundamental limits of parameter estimation and target detection and the connection between sensing and communication. He is the first author of two monographs, Spatial Information Theory and Sensing Information Theory, and a coauthor of several other monographs. He has authored or coauthored more than 300 journal and conference papers and holds more than 40 granted or pending patents. One of his papers was selected as one of China’s Most Influential International Academic Papers. He serves on the Academic Committee of the National Mobile Communications Research Laboratory at Southeast University, the Academic Committee of the Frontiers Science Center for Mobile Information Communication and Security of the Ministry of Education, China, and the Technical Committee on Communication Theory and Signal Processing, China Institute of Communications. 
\end{IEEEbiographynophoto}

\begin{IEEEbiographynophoto}{Nan Wang}
		received the Ph.D. degree in communication and information system from the Nanjing University of Aeronautics and Astronautics, Nanjing, China, in 2025. She is currently a lecturer with the Shenyang Aerospace University, Shenyang, China. Her research interests include information theory for sensing, spatial information theory, channel coding, and radar signal processing.
\end{IEEEbiographynophoto}

\begin{IEEEbiographynophoto}{Han Zhang}
	received the Ph.D. degree in communication and information system from the Nanjing University of Aeronautics and Astronautics, Nanjing, China, in 2023. He is currently a lecturer with the Nanjing University of Posts and Telecommunications, Nanjing, China. His research interests include the broad areas of radar signal processing and information theory.
\end{IEEEbiographynophoto}

\end{document}